\documentstyle[12pt]{article}
\newcommand{\be}{\begin{equation}}
\newcommand{\ee}{\end{equation}}
\newcommand{\ba}{\begin{eqnarray}}
\newcommand{\ea}{\end{eqnarray}}

\title{Baryogenesis from Domain Walls in the
Next-to-Minimal Supersymmetric
Standard Model}
\author{S.~A.~Abel$^{1}$ and P.~L.~White$^{2}$
\\$^{1}$Rutherford Appleton Laboratory
\\Chilton, Didcot
\\Oxon OX11 0QX
\\England
\\ \vspace{0.3cm}
\\$^{2}$Theoretical Physics
\\University of Oxford
\\1 Keble Road
\\Oxford OX1 3NP
\\England}

\begin{document}

\hfill OUTP-9517P, hep-ph/9505241 \\

\begin{center}
{\Large \bf Baryogenesis from Domain Walls in the
Next-to-Minimal Supersymmetric
Standard Model}
\bigskip \\
{S.~A.~Abel$^{1}$ and P.~L.~White$^{2}$
\\$^{1}$Rutherford Appleton Laboratory
\\Chilton, Didcot
\\Oxon OX11 0QX
\\England
\\
\vspace{0.3cm}
$^{2}$Theoretical Physics
\\University of Oxford
\\1 Keble Road
\\Oxford OX1 3NP
\\England}
\end{center}

\small

%\maketitle

\title{Abstract}

\begin{abstract}
We consider the production of baryon number from collapsing domain
walls, and in particular examine the magnitude of  CP violation which
is required in such schemes. Taking the conventional solution to the
domain wall problem in the Next-to-Minimal Supersymmetric Standard
Model as an example, we show that the observed baryon assymmetry of the
universe may have been generated, even if the initial {\em explicit} CP
violation in the Lagrangian were so small ({\em i.e.} gravitational)
that it could never be experimentally  detected. This is possible by
having the {\em explicit} CP violation affect the way in which the
walls collapse, rather than be responsible for the  generation of
baryon number directly. Net baryon number is created at the domain
walls by the spontaneous breaking of CP.
\end{abstract}

\pagebreak

\section{Introduction}

In models of baryogenesis at or close to the electroweak phase transition,
one of the three Sakharov conditions \cite{SAK}
(no thermal equilibrium) is fulfilled by
the phase transition itself, while the remaining two conditions (B, C and CP
violation) are provided by sphaleron  mediated processes and some extension to
the Higgs sector usually involving an explicit CP violating phase
\cite{CK88,ACW93}. So far, the study of baryon number production
involving topological defects has mainly
addressed the first of these conditions.
That is, it has been shown that  the departure from thermal equilibrium may be
provided by the collapse of  cosmic strings or domain walls \cite{anne,lew}. In
this paper we shall address the second. It is generally assumed that there must
be an explicit CP violating extension to the Higgs sector similar in magnitude
to that  required for electro-weak baryogenesis, in order to bias the
production of baryons.  Such CP violating terms may eventually be detected
through their contribution to the neutron electric dipole moment for example
\cite{edm}.  Here we shall show that this is not a necessary condition
for baryogenesis from topological defects. In fact, provided that there is
{\em spontaneous}  breaking of CP when the domain walls form \cite{lee},
it is possible to
generate the observed baryon  asymmetry with additional {\em explicit} CP
violating terms which are gravitationally  suppressed, and which will therefore
never be detected. We stress the difference between `spontaneous' violation
of CP which is responsible for the local production baryon number at the
domain walls, and the `explicit' CP violation which is needed in order to have
a global excess of baryons. In this respect, our picture is similar to that
proposed in Ref.\cite{comelli}.

Our argument can be summarised as follows.
Because the domain walls in question result from a breaking of CP, any
particular wall is not CP invariant. Global CP invariance is provided by the
fact that there exist different types of wall which are CP conjugates of each
other. The mechanism which is invoked in order to remove the walls need not
involve large terms in the Higgs potential. In fact for walls in which the
higgs VEV, $v$, is of the order the electroweak scale (or larger), if the
degeneracy in the minima is broken by gravitational couplings of order
$v^5/M_{pl}$, the walls will disappear well before the onset of
nucleosynthesis \cite{zel}. Since individual walls are not CP invariant,
it is possible to generate a sufficient baryon number with explicit
CP violation of order $v^5/M_{pl}$.

We shall demonstrate this using the next-to-minimal supersymmetric
standard model (NMSSM).  It should be borne in mind however that our
discussion applies to any model in which the spontaneous breaking of CP
produces  domain walls.  By choosing an example with three phases we
are perhaps making things more  difficult for ourselves, since models
with more than two phases  have their own special problems (some of
which will be addressed in Ref.\cite{forthcoming}). However it is
interesting that a case with the required properties exists already in
the literature.

The NMSSM \cite{NMSSM} is an extension of the usual minimal
supersymmetric standard model (MSSM) \cite{MSSM}, in which the usual
two Higgs doublets $H_1$ and $H_2$ which are necessary to give masses
to the up and down type quarks are supplemented by a singlet Higgs
superfield $N$. The usual $\mu$ term in the lagrangian, $\mu H_1H_2$,
is then eliminated by invoking a $Z_3$ symmetry under which every
chiral superfield $\Phi$ transforms as $\Phi\to e^{2\pi i/3}\Phi$. The
allowed terms in the superpotential are then $\lambda
NH_1H_2-\frac{k}{3}N^3$, in addition to the usual fermion mass
generating Yukawa terms, while the Higgs part of the soft supersymmetry
breaking potential is extended by the inclusion of two more extra
trilinear soft terms $A_\lambda$ and $A_k$ in place of the MSSM term
$B\mu H_1H_2$ to become
\ba
V_{soft}^{Higgs}=
- \lambda A_{\lambda}(NH_1H_2 + h.c.)
- \frac{k}{3} A_k (N^3 + h.c.)
\nonumber\\
+ m^2_{H_1}\vert H_1\vert^2
+ m^2_{H_2}\vert H_2\vert^2
+ m^2_{N}\vert N\vert^2
\ea
where $H_1H_2=H_1^0H_2^0-H^-H^+$, and we shall hereafter drop the 0
index for neutral Higgses.

The primary motivations for the NMSSM are the elimination (or at least
reparametrisation) of the $\mu$ problem, which is that it is not clear
what could be the origin of a $\mu$ parameter in $\mu H_1H_2$ which is
of order the electroweak scale; that it allows the evasion of the usual
MSSM Higgs mass bounds \cite{NMSSMbound}; and the fact that this
relatively minor alteration to the model gives an extremely rich and
complex Higgs and neutralino phenomenology which can be significantly
different from that in the MSSM \cite{NMSSMphen}.

When electroweak symmetry breaking occurs, the three neutral CP-even
Higgs scalars acquire VEVs. Using the parameters
\ba
H_1 &=& \rho_1 e^{i\theta_1}\nonumber\\
H_2 &=& \rho_2 e^{i\theta_2}\nonumber\\
N  &=& \rho_x e^{i\theta_x}
\ea
it can be shown that any true minimum of the potential does not violate
CP in the sense that the VEVs can always be made real by an appropriate
field redefinition, up to the existence of three degenerate vacua
related to each other by $Z_3$ transformations \cite{romao}, and hence
we have minima with $\theta_1=\theta_2=\theta_x=\frac{2\pi ni}{3}$ for
integer n, \footnote{We can use weak hypercharge to select any phase
for $\theta_1-\theta_2$.} and with $\rho_i=\nu_i$, $\rho_x=x$. Note
that although we have imposed that the $\nu_i$ be real, one or two of
them may still be negative. We shall refer to the three minima
\ba
\theta_1+\theta_2&=&0, 2\pi /3, 4 \pi/3\nonumber\\
\theta_x&=&0, 4\pi /3, 2 \pi/3,
\ea
as $A$, $B$, $C$ respectively, and for convenience will assume that the
evolution of the universe will ultimately end with phase A dominating.

After the electroweak phase transition the universe will be divided up
into regions of different minima separated by domain walls.  In each of
the three degenerate minima, there is an operation which performs a CP
transformation in the effective low energy theorem, and  which maps the
vacuum into itself. If we define the $Z_3$ operation to be
$Z_3:A\rightarrow B$, $Z_3:B\rightarrow C$, and $Z_3:C\rightarrow A$,
then the transformations are
\ba
\mbox{CP}_A &=& \mbox{CP} \nonumber\\
\mbox{CP}_B &=& \mbox{CP} Z_3  \nonumber\\
\mbox{CP}_C &=& \mbox{CP} Z_3^2,
\ea
where here, CP is the transformation in the full theory. In each minimum,
the two false vacua are CP conjugates of each other. Alternatively,
we could have performed a
field redefinition such that the true minimum (here A) is CP
invariant.

\section{Domain Walls in the NMSSM}

Domain walls are one of the simplest types of topological defects
\cite{zel}, and form whenever the theory in question has a discrete
number of degenerate vacua, usually due to the spontaneous breaking of
a discrete symmetry. The simplest example occurs in the case of a real
scalar field $\phi$ with potential $V=\lambda (\phi^2-\nu^2)^2$. This
potential clearly has two degenerate minima with $\phi=\pm\nu$ as a
result of the $Z_2$ symmetry $\phi\to-\phi$. If we look for time
independent solutions of the field equations which are translation
invariant in two of the three space dimensions, and which obey
$\phi(z=-\infty)=-\nu$ and $\phi(z=\infty)=\nu$ we find a solution
$\phi=\tanh(z/\delta z)$. This is a domain wall, whose thickness
$\delta z$ is given by $\delta z=(\sqrt{2\lambda}\nu)^{-1}$, and which
has a surface energy $\sigma$ given by $3\sigma=4\sqrt{2\lambda}\nu^3$.
In more complicated models, it is no longer possible to solve the field
equations analytically, but it is straightforward to solve them
numerically. We find that $\delta z\sim\nu^{-1}$ and $\sigma\sim\nu^3$
as before, where $\nu$ is now some typical VEV of one of the fields,
and the structure is in general similar.

Turning specifically to the NMSSM, the potential for the neutral
scalars takes the form (at tree-level)
\ba
V=\frac{(g_1^2+g_2^2)}{8}(\vert H_1\vert^2 - \vert H_2\vert^2)^2
+ \lambda^2 \vert N\vert^2(\vert H_1\vert^2 + \vert H_2\vert^2)
\nonumber\\
+ \lambda^2 \vert H_1\vert^2 \vert H_2\vert^2
- \lambda k (\bar N^2H_1H_2+N^2\bar H_1\bar H_2) \nonumber\\
+ k^2 \vert N\vert^4
+ m^2_{H_1}\vert H_1\vert^2
+ m^2_{H_2}\vert H_2\vert^2
+ m^2_{N}\vert N\vert^2 \nonumber\\
- \lambda A_{\lambda}(NH_1H_2 + \bar N\bar H_1\bar H_2)
- \frac{k}{3} A_k (N^3 + \bar N^3)
\ea
With real VEVs $<\rho_1>=\nu_1$, $<\rho_2>=\nu_2$, $<\rho_x>=x$ our
inputs are then $\tan\beta=\frac{\nu_2}{\nu_1}$, $r=\frac{x}{\nu}$,
$\lambda$, $k$, $A_{\lambda}$, $A_k$, while $\nu^2=\nu_1^2+\nu_2^2$ is
derived from the requirement that we have the correct Z mass. We choose
to specify the VEVs as input parameters rather than the masses
appearing in the potential for convenience, since we may immediately
calculate the soft masses $m^2_{H_1}$, $m^2_{H_2}$, $m^2_{N}$ from the
VEVs by using the minimisation conditions. Of course, this model
typically has several different minima, usually including for example
minima with only one of the three VEVs non-zero, and in order to study
the vacuum structure it is necessary to find all of them to ensure that
the minimum which we are analysing is indeed the deepest one.

Let us now turn to domain wall solutions of the field equations. These
reduce to the six equations
\be
\frac{d\phi_i}{dz}+\frac{1}{2}\frac{\partial V}{\partial\phi_i}=0
\ee
where $\phi_i$ is the real or imaginary part of one or other of the
three scalar Higgs fields. We may then impose the boundary conditions
that $(H_1,H_2,N)$ are $(\nu_1,\nu_2,x)$ at $z=-\infty$ and
$(\nu_1e^{2\pi i/3},\nu_2e^{2\pi i/3},xe^{2\pi i/3})$ at $z=\infty$.
It is straightforward to find solutions to such equations numerically,
and by appropriate $Z_3$ field redefinitions it is clear that
walls with the same structure exist between any
two pairs of vacua.

A typical solution is shown in Figure 1, where we show the absolute
values, phases, and energy density as a function of z in the wall
region. The input parameters are  $\lambda=k=0.2$,
$A_{\lambda}=A_k=100$GeV, $\tan\beta=r=2$. Here the total surface
energy density of the wall is  $7.1\times 10^6\hbox{GeV}^3$ after we
have subtracted the vacuum energy density, while the wall thickness is
around $0.02\hbox{GeV}^{-1}$, in reasonable agreement with the
approximate arguments given above. It should be noted that even in the
centre of the wall the VEVs are not zero, and so electroweak symmetry
is not restored.

In fact, as the parameters are varied a very wide range of different
behaviours and structures for the wall can be seen, with typically (for
$\tan\beta>1$ and $r>1$) $\rho_x$ remaining large over much of the
region, while $\rho_2>0$ always but may become quite small near the
centre of the wall. The phase behaviour shown in Figure 1b, where the
$U(1)_Y$ phase $\theta_1-\theta_2$ goes from $2\pi$ to 0 continuously
across the range is not universal but is typical. Unlike that for all
the other variables, most of this change in $\theta_1-\theta_2$ is
outside the wall region, but we have explicitly checked that changing
the size of the box does not have any significant impact on the total
energy or the shape of the field configuration.

An example of a set of parameters for which electroweak symmetry is
virtually restored is shown in Figure 2. Here $\lambda=k=0.1$,
$A_{\lambda}=A_k=250$GeV, $\tan\beta=2$, $r=5$. The total wall energy
is $2.2\times 10^7\hbox{GeV}^3$, rather higher than before because the
singlet VEV is larger, while the wall is now slightly wider. Although
only $\rho_1$ is ever zero inside the wall, $\rho_2$ falls to under
3GeV, and is less than 10GeV for a region of width $\sim
0.04\hbox{GeV}^{-1}$.

Of course, these are just two of a multitude of possible sets of
parameters, each of which will give a wall with possibly very different
characteristics. We also remark that there can be more than one
solution for a given set of parameters, although for those shown there
are no other wall solutions with higher surface energy. Since the VEVs
of the  higgs fields do not go through the origin, there is the
possibility  for `double' (and also triple in this case) wall systems
with the same phase on either side to form in  the manner described in
Ref.\cite{dvali}. We shall assume that this does  not occur, or at
least that if it does the double walls are unstable to the formation
of holes by quantum tunneling, which then expand under the surface
tension  destroying the wall. Our primary conclusion must be that for
at least some sets of parameters, the domain walls possess exactly the
properties which we will require in order to have them driving
baryogenesis.

\section{Baryogenesis from $Z_3$ Wall Networks}

Having established this fact, let us go on to examine the possibilities
for baryogenesis. After the phase transition we have an `emulsion' of
three phases separated  by highly convoluted domain walls. CP is also
spontaneously broken by the phase transition, but as yet there is no
{\em explicit} CP violation.  In fact, as we have seen, when going
through a  domain wall from $A\rightarrow C$, $B\rightarrow A$ or
$C\rightarrow B$, the phase changes of the Higgs fields are equal and
opposite to those occurring when going from $A\rightarrow B$,
$B\rightarrow C$ or $C\rightarrow A$. We shall refer to  these two
types of transition as `positive' and `negative' respectively. The
walls are not invariant under CP, and inside them the electroweak
symmetry is restored if the vacuum expectation values of the Higgs
fields vanishes (which, as we have seen, may or may not be the case
depending on the details of the Higgs sector). Thus we shall assume
that baryon number  violating transitions will be in equilibrium in
these regions, at plasma temperatures close to the phase transition.
As domain walls move through space, the time-dependent change of phase
of the Higgs fields occurring inside the walls will give rise to a
non-zero chemical potential for baryon number and baryons will
therefore be created.

Cosmology dictates that there is some mechanism which removes the walls
and one suggestion, originally by Zel'dovich {\em et al} \cite{zel},
is that the degeneracy of the vacua may be slightly broken, eventually
leading to the dominance of the true vacuum.  This point of view was
recently supported by Rai and Senjanovic \cite{rai}, who argue that
gravitational interactions may explicitly break the discrete symmetries
causing a slight non-degenaracy in the minima of the Higgs of order
$\epsilon \sim v^5/M_{pl}$  (where $v$ is a generic Higgs VEV of order
$M_W$ in this example). This suggestion was applied to this particular
model in the context of string theories by Ellis {\em et al}
\cite{ellis}.

We should point out two possible problems with this  solution to the
domain wall problem for this particular model. The first is the problem
of destablising divergences which may generate a large VEV for the
singlet, and so destroy the solution which supersymmetry provides for
the hierachy problem \cite{bagger}. This is a potential fault in any
model which includes gauge singlets. It is not clear which operators
may be generated at the Planck scale or with what coefficients, but we
note that the CP-violating gravitationally-suppressed operators which
are necessary to remove domain walls do not in themselves generate such
large singlet VEVs. Connected with this problem is the fact that if we
break the $Z_3$ symmetry even by gravitational terms, we reintroduce
the $\mu$-problem since without the $Z_3$ symmetry there is nothing to
prevent $\mu$ becoming large. These points detract from the NMSSM but
are unavoidable; unless we are prepared to complicate the model by
invoking inflation with reheating to a temperature less than the weak
scale (and probably the Affleck-Dine mechanism for baryogenesis), we
must certainly break the $Z_3$ symmetry explicitly.  These are problems
for the NMSSM as a whole and are secondary to our present more general
aim of showing that domain walls can induce baryogenesis with small CP
violation. We will not discuss them further, but will simply bear in
mind that a full resolution of these problems of the NMSSM seems  to
require a greater understanding of the structure at the Planck scale.

The removal of the false vacua (and therefore the domain walls)
proceeds as follows. For friction-free motion, the typical curvature
scale, $R$,  of the wall structure evolves roughly as the time for
models with $Z_N$  symmetry. Since we are not interested in the
precise power law behaviour of the  curvature scale, we shall neglect
the conformal stretching due to the  expansion of the universe. For
detailed discussions of these points see Refs.\cite{kawano,press}. We
shall also neglect the effects of friction  on the motion of the walls.
In fact this may be important at lower temperatures for domain walls
associated with higgs fields. This is due to the  walls' interaction
with particles in the plasma, most  importantly the bottom quarks,
which are reflected off them with  probability proportional to
$m_b^2/p^2$ where $p$ is the particle's momentum.  Thus friction is
unimportant for temperatures between $E_W$ and $~10$ GeV.  When the
motion of the walls is friction dominated, they reach a terminal
velocity determined by their curvature and by the density of the
plasma. The typical curvature scale of the walls then increases as
$t^{1/2}$ rather  than $t$ \cite{lifshitz}.  These points will be
discussed in detail in \cite{forthcoming}.

Once the curvature scale has exceeded a critical value, {\em i.e.}
when the pressure dominates over the tension, $\epsilon > \sigma/R$
where $\sigma$ is the surface energy density, the domains of true vacuum
begin to dominate and expand into the two domains of false vacuum.
However, since CP is only broken spontaneously,
any mechanism which removes the walls generates as much matter as anti-matter.
In this case, the true vacuum, $A$, invades an equal area
of $B$ and $C$ when it finally dominates
($B$ and $C$ must be degenerate if CP is not explicitly broken),
and the production of baryons
from negative walls exactly cancels that from positive ones.

Spontaneous CP violation {\em per se} is therefore not enough to
generate a  net  baryon number. What is also required is some
additional {\em explicit} CP  violation in the Lagrangian, and this is
where this paper differs from previous  discussions. Previously
attention has almost always been focussed on the  biasing of the baryon
number production directly at the collapsing domain walls (an exception
being the scenario examined in Ref.\cite{comelli}, which  bears some
resemblances to this picture).  Thus any explicit CP violation  that
was added to the Lagrangian was incorporated linearly into the
production of baryon number. The resultant models required relatively
large  CP violating phases in the higgs sector.

However, even tiny (of order $v^5/M_{pl}$) CP violating terms
will clearly effect the way the domain walls collapse, and,
as argued in Ref.\cite{rai}, there is no reason why
gravitational terms that break the $Z_3$ should not also break
CP. What we propose
therefore, is that no two of the vacua are degenerate, so that $C$ has a
higher vacuum energy than $B$, which has a higher vacuum energy than $A$.
As the walls collapse therefore, $A$ domains will invade
both $B$ and $C$, $B$ domains will invade $C$ but be invaded by $A$, and
$C$ will be invaded by both $A$ and $B$. In order to show that this can
generate a significant baryon asymmetry, consider the extreme case, in
which the $C$ phase
has `much' higher vacuum energy than $B$. Then the first pressure
driven process to operate as the scale of the wall network increases
is the collapse of $C$ domains to be replaced by
$A$ and $B$. Both positive and negative walls will
quickly accelerate to the speed of light, and the net baryon number
generated will be close to zero. Now there will be only $A$ and $B$
phases left.
The remaining $B$ is finally removed when the non-degeneracy in $A$ and $B$
vacua becomes dominant. But the only walls which can do this are
the positive $B\rightarrow A$ ones, and so there is clearly the
potential for generating net baryons.

\section{Simulation of the $Z_3$ Wall Networks}

The number which we need in order to be able to estimate the baryon
number is the average number of positive and negative walls which pass
through a given point  during the whole process. In order to show that this
number can be close to one,
we have simulated a  $Z_3$ domain wall network evolving in 2
dimensions, in Minkowski space ({\em i.e.} neglecting the conformal
stretching).  We did this  following Kawano \cite{kawano}. First we
begin with an arbitrary distribution  of the three phases. The
probability for each phase is $P_A=P_B=P_C=0.33$  which interestingly
is barely enough for them to percolate in three dimensions.
Simulations on a cubic lattice give the percolation threshold to be
$0.31$ and  simulations in continuum percolation theories give
$0.295\pm 0.02$ \cite{percol}.  Thus the structure is expected to be
tenuous and highly convoluted ({\em i.e.} `spaghetti'--like).  The
walls are then divided into small lengths and released from the (in
this case square) lattice. The  evolution at each time step is determined
by applying the equations of motion  locally, taking the mass
(proportional to the length) of the walls to be  concentrated at the
vertices between straight sections. In this we differ slightly from
Kawano, who calculated the local curvature, since this  enabled us to
treat vertices with two and three walls attached on the same footing.
In addition we did not include toroidal boundary conditions but let the
 ends of the walls slide along the edge of the box. We therefore do not
expect our results to be accurate when the curvature scale is of the
same order as the size of the box.  Our basic unit for the simulation
is shown in figure 3. A typical point, $r_0$, is connected to up to
three other points.  Each line has a perpendicular vector
$\epsilon_{ij}$  associated with it which describes the magnitude and
direction of the pressure acting on it.The rest mass of the vertex is
given by half the sum of the lengths multiplied by the surface density
$\sigma$;
\be
m_0=\frac{\sigma}{2}\sum_{i}|r_i-r_0|.
\ee
The force is given by $-\nabla E$ at the vertex
\be
\frac{\partial p_0}{\partial s}
=\sum_{i}\left( \gamma\sigma\frac{r_i-r_0}{|r_i-r_0|}
        +\epsilon_{0i}|r_i-r_0|\right),
\ee
where $\gamma=(1-\dot{r}_0^2)^{-1/2}$ and $s$ is proper time,
so that the acceleration is given by
\be
\label{accn}
\frac{d^2 r_0}{dt^2}
=\sum_{i}\left( 2 \gamma^{-2}\frac{r_i-r_0}{|r_i-r_0|}
        +\frac{\epsilon_{0i}\gamma^{-3}}{\sigma}|r_i-r_0|\right)/
\sum_{j}|r_j-r_0|.
\ee
It is easy to verify that the continuous case is recovered in the
limit as the size of the straight sections goes to zero.
For example, consider the polygon made of $N$ equal straight lengths, whose
vertices are at $r$ from the origin, and whose internal vacuum energy
density is $\epsilon$. The equation of motion above leads to
\be
\frac{d^2 r}{dt^2}
= -\frac{(1-\dot{r}^2)}{r} -
        \frac{\epsilon\gamma^{-3}}{\sigma}\cos\frac{\pi}{N},
\ee
which is that of a cylinder of radius $r$ in the limit
$N\rightarrow\infty$ \cite{widrow}.
It is convenient to scale everything in terms of the initial
curvature scale $R_0$, so that $\rho_i=r_i/R_0$, and $\tau=t/R_0$,
so that eq.(\ref{accn}) becomes,
\be
\label{accn2}
\frac{d^2 \rho_0}{dt^2}
=\sum_{i}\left( 2 \gamma^{-2}\frac{\rho_i-\rho_0}{|\rho_i-\rho_0|}
        +\frac{\epsilon_{0i}R_0\gamma^{-3}}{\sigma}|\rho_i-\rho_0|\right)/
\sum_{j}|\rho_j-\rho_0|.
\ee
The only free parameters in the simulation are therefore the
two pressure variables, $\epsilon_B R_0/\sigma$ and $\epsilon_C R_0/\sigma$.
All our results are presented with $R_0$ normalised to 1cm.

In figure 4 we can see how the network behaves without the effects of pressure.
The scale of the structure evolves  at roughly the speed of light (with about
one wall per horizon)  growing proportionally to the  time. (This case,
together with more general $Z_N$ cases has been  examined in more detail by
Press {\em et al} \cite{press}.) The important point here is that (as remarked
upon in Ref.\cite{lifshitz}),  without pressure, the evolution  of the walls is
mostly a question of topology. Those regions which are connected to two or four
external walls tends to collapse, while those which are connected to eight or
more external legs expand, due to the tension of the external legs pulling
outwards.  One can see this by considering any three leg vertex. A three leg
vertex minimises its wall energy by trying to adopt a position in which the
angles between the legs are equal and $120^o$. Squares with four  external legs
do this by the vertices falling inwards trying to increase the internal $90^o$
angle. Octagons with eight external legs expand, trying to decrease a $135^o$
angle. Hexagonal structures ({\em i.e.} honeycombs) are stable. In fact for a
general $N$ sided polygon with $N$ external legs, the equation of motion is
easily found to be
\be
\label{poly}
\frac{\partial^2}{\partial \tau^2} r=
-\frac{1}{r\gamma^2} \left( 1 -
  \frac{1}{2\gamma\sin\left(\frac{\pi}{N}\right)}\right)
+\frac{\epsilon R_0}{\sigma\gamma^3}
\ee
where we have normalised $r=R/R_0$, where $R$ is the perpendicular
distance from the centre to the edges of the polygon, and $\tau=t/R_0$.
$\epsilon$ is the difference in vacuum energy between the inside and
outside of the polygon, and $R_0$ is its initial size.

We now introduce pressure by switching on the $\epsilon$ above.
This becomes dominant over the tension when
\be
\left|\frac{\epsilon R}{\sigma}\right|\sim 1
\ee
which for typical values of vacuum expectation values  happens for
$1cm<R_0<1m$ in cases where $\epsilon$ is induced gravitationally.
The evolution of the system with pressure is shown in figure 5,  where
we have taken $\frac{\epsilon R_0}{\sigma} = 0,0.25,0.5$ for the phases
$A,B,C$ respectively. As mentioned earlier, the evolution of the
network in terms of $r=R/R_0$ is the same for constant $\epsilon R_0$.
As in the $Z_2$ case, larger  structures are affected much more by the
pressure. Since the structure  of the walls is always increasing, once
the pressure becomes dominant,  collapse happens very quickly (between
$10^{-10}$ and $10^{-8}$ seconds). Thus provided that the walls do not
dominate in density before they collapse, and also that the entropy
released into the plasma is properly thermalised (both of which we
shall assume), there is no danger of disturbing nucleosynthesis which
begins at $\sim 1$ second.

The ratio of area cleared by positive walls to the total area we find to be
$0.6$, so that definining $\kappa_{BG}$ as
\be
\kappa_{BG}=\frac{\hbox{area of positive transitions -
  area of negative transitions}}
{\hbox{total area}}
\ee
the global production of baryons is $\kappa_{BG}\approx 0.2$ of what it
would be for maximal CP violation. For values of $\epsilon_c$ much
larger than this, $\kappa_{BG}$ rapidly approaches 1.

\section{Discussion.}

 From this point, the analysis closely follows that in Ref.\cite{anne},
and of course the same caveats apply. That is, we assume that the  wall
thickness is large enough to allow anomalous processes to occur. These
may take the form of short range interactions of typical size
$(g^2T)^{-1}$, where the electroweak symmetry is completely restored, or
if the temperature is close to the  phase transition one would expect
sphaleron-like configurations straddling  the domain wall to be
possible. (Ideally one would like to be able to find these by
constructing a non-contractible loop around the domain wall
background.) Assuming that the sphaleron rate inside the walls is
\be
\Gamma_B\sim \kappa(\alpha_W T)^4
\ee
the final production of baryons is given by
\be
\frac{n_B}{s}\approx \kappa
\kappa_{BG}g_*^{-1}\alpha_W^4\Delta(\theta_1+\theta_2)
\ee
where $g_*$ is the number of relativistic degrees of freedom
contributing to the entropy, and $\Delta(\theta_1+\theta_2)$ is the
phase change of the higgs fields which is orthogonal to the Goldstone
mode. This factor is to be multiplied by a coefficent of order 1 (see
Ref.\cite{ACW93}).  Note that in contrast to `spontaneous baryogenesis'
scenarios in which there is  a possible suppression by a factor
$m_t^2/T^2$ \cite{dine}, here there is no suppression,  because the
relevant {\em equilibrium densities} to use are the ones obtaining
outside the  domain wall where $m_t$ has its physical value. Notice
that the production of baryon number is due to the translation of the
walls, which contrasts with the case of electroweak strings, in which
it is due to a decrease in the total volume covered by strings as they
collapse; because of this there is no volume suppression factor $(SF)$.
The change in higgs phase is
\be
\Delta(\theta_1+\theta_2)\sim 2 \pi/3,
\ee
so that
\be
\frac{n_B}{s}\approx 10^{-8}\kappa\kappa_{BG}g_*^{-1}.
\ee

Bearing in mind our earlier discussion, there is the possibility of a
much larger biasing of the potential. We note that this mechanism works
for more general $\epsilon$, provided firstly that the amount of
explicit CP violation is of the same order as the explicit $Z_3$
breaking, and secondly that they are both not so large that the walls
collapse immediately on forming. Thus we are in the novel position of
being able to place (albeit extremely weak) lower and upper bounds on
the amount of explicit CP violation allowed.  The mechanism works only
when the scale at which  the pressure dominates is larger than the
size of the protodomains ($\approx (g^2_2 T)^{-1}$) during the phase
transition which gives,
\be
\epsilon \stackrel{<}{\sim} g_2^2 T_c \sigma .
\ee
For typical values of $\sigma$ we find,
\be
\epsilon \stackrel{<}{\sim} 10^7 \mbox{GeV}^4
\ee
which not surprisingly is just less than $M_W^4$.  In addition we
require that the temperature be close to the weak scale, for the
anomalous processes to be operative inside the wall.  However the
domain sizes grow at speeds comparable to the speed of light.  FRW
cosmology gives $t=$2.42~$g_*^{-1/2} (T/\mbox{MeV})^{-2}$ secs.  For
$T\sim M_W$  we find $t\sim 10^{-10}$ secs. Thus it is possible for 1
cm size structures to grow at temperatures close to the weak scale,
implying that even CP violation induced by gravity could be the driving
force behind baryon production for weak scale phase transitions.

In fact the following exercise is instructive. Suppose that the anomalous
processes are effective down to a temperature $T_*<T_c$, and that the
pressure and surface energy terms are given by
\be
\epsilon = v^5/M \mbox{ ; } \sigma=v^3
\ee
where $v={\cal O}( M_W)$ is the VEV of the higgs fields, and $M$ is the mass
scale of the physics which is responsible for the CP violation.
Suppose also that the curvature scale increases at some sizeable fraction,
$\beta$, of the speed of light, $R=\beta (t-t_c)$. Then in order for
this mechanism to work, we require that the pressure is dominant over the
surface tension for $t=t_*$,
\be
\epsilon > \sigma / \beta(t_*-t_c).
\ee
This gives an upper bound on $M$,
\be
M\stackrel{<}{\sim}M_{pl} \left( 0.3 \beta g_*^{-1/2}
\left(\frac{v}{T_*}\right)^2
\left( 1-T_*^2/T_c^2 \right)\right).
\ee
So unless $T_*$ is extremely close to $T_c$, gravitational couplings could
be responsible for this mechanism, giving a lower bound on $\epsilon$ of,
\be
\epsilon \stackrel{>}{\sim} 10^{-8} \mbox{GeV}^4.
\ee
When this bound is saturated, on dimensional grounds one expects the
contribution to the electric dipole moment of the explicit CP violation,
to be of the order of $\delta d_n < 10^{-42}e\mbox{cm}$ \cite{edm}.

Finally, we note that this mechanism is possible for
any model with a spontaneous CP breaking transition occuring at an
energy scale, $v$, which is higher than the electroweak scale, provided that
some domain walls remain at the time of the electroweak transition.
In this case anomalous processes are guaranteed to be in equilibrium
when the wall network collapses. The same considerations apply here.
That is
\be
g^2 v^4 \stackrel{>}{\sim} \epsilon \stackrel{>}{\sim} 10^{-8} \mbox{GeV}^4.
\ee
In this case the lower bound is {\em less} than what would be expected  to be
induced by gravity, since we require simply that the domain walls  collapse
before the electroweak phase transition whilst anomalous processes are still in
equilibrium.

\vskip 0.25in

{\bf\Large Acknowledgement} \hspace{0.3cm}

We would like to thank J.~Ellis, G.~G.~Ross, M.~Shaposhnikov, and
especially S.~Sarkar for conversations.

\pagebreak

\newpage
{\center{\bf Figure Captions}}

{\bf Figure 1a}\\
Absolute magnitures of the fields as a function of position
for the parameters $\lambda=k=0.2$, $A_{\lambda}=A_k=100$GeV,
$\tan\beta=2$, $r=2$. The three lines show, from top to bottom,
$\rho_3$, $\rho_2$, $\rho_1$.

{\bf Figure 1b}\\
Phases of the fields as a function of position for the same parameters
as Figure 1a. The three lines show  $\theta_+=\theta_1+\theta_2$ (solid
lines) , $\theta_-=\theta_1-\theta_2$ (long dashed lines), $\theta_x$
(short dashed lines).

{\bf Figure 1c}\\
Surface energy density of the wall as a function of position for the
same parameters as Figure 1a.

{\bf Figure 2a}\\
Absolute magnitures of the fields as a function of position
for the parameters $\lambda=k=0.1$, $A_{\lambda}=A_k=250$GeV,
$\tan\beta=2$, $r=5$. The three lines show, from top to bottom,
$\rho_3$, $\rho_2$, $\rho_1$.

{\bf Figure 2b}\\
Phases of the fields as a function of position for the same parameters
as Figure 2a. The three lines show  $\theta_+=\theta_1+\theta_2$ (solid
lines) , $\theta_-=\theta_1-\theta_2$ (long dashed lines), $\theta_x$
(short dashed lines).

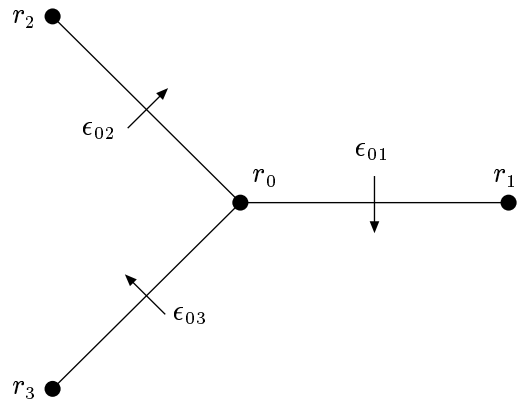
{\bf Figure 3}\\
Basic unit for wall simulation.

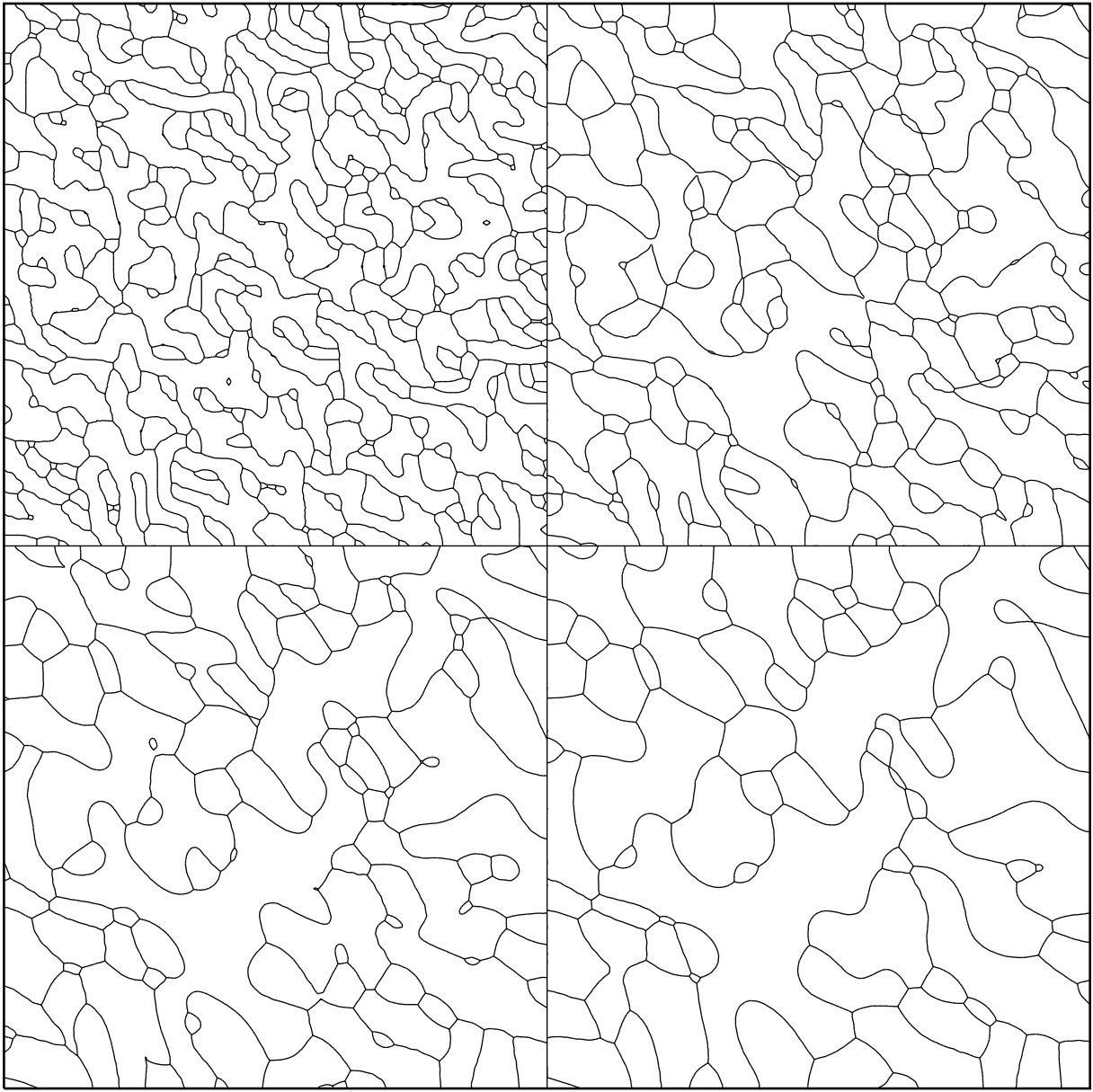
{\bf Figure 4}\\
Wall evolution without pressure. The four time-slices shown have time
0.5 $10^{-10}$s, 1.5 $10^{-10}$s, 2.5 $10^{-10}$s, 3.75 $10^{-10}$s,
for upper left, upper right, lower left, lower right respectively.
$\kappa_{BG}$ is less than 0.01 always.

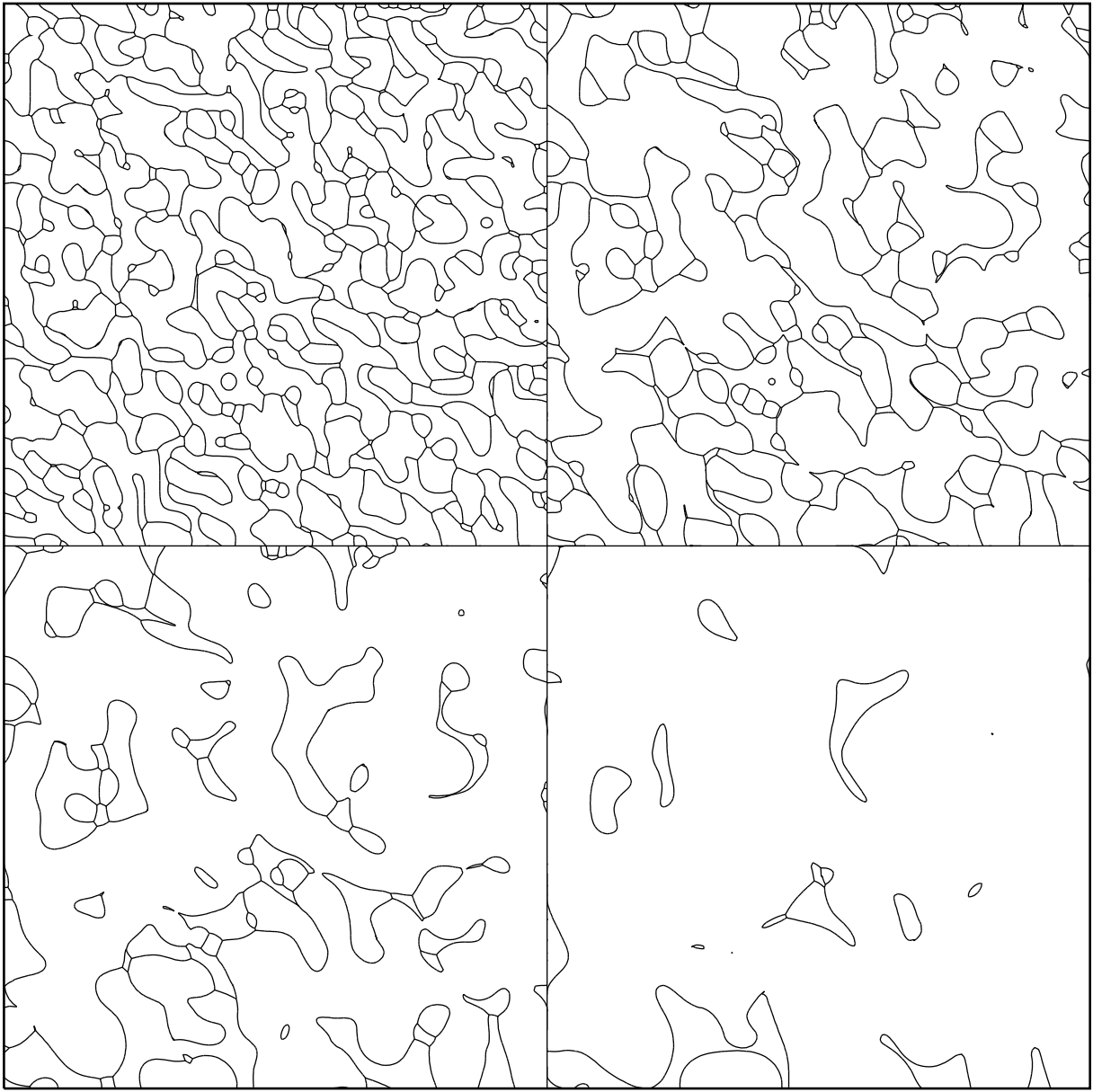
{\bf Figure 5}\\
Wall evolution without pressure. The four time-slices shown have time
0.6 $10^{-10}$s, 1.5 $10^{-10}$s, 2.4 $10^{-10}$s, 3.75 $10^{-10}$s,
for upper left, upper right, lower left, lower right respectively.
$\kappa_{BG}$ is -0.009, -0.004, 0.023, 0.104. Final value of
$\kappa_{BG}$ after all walls disappeared was around 0.2.

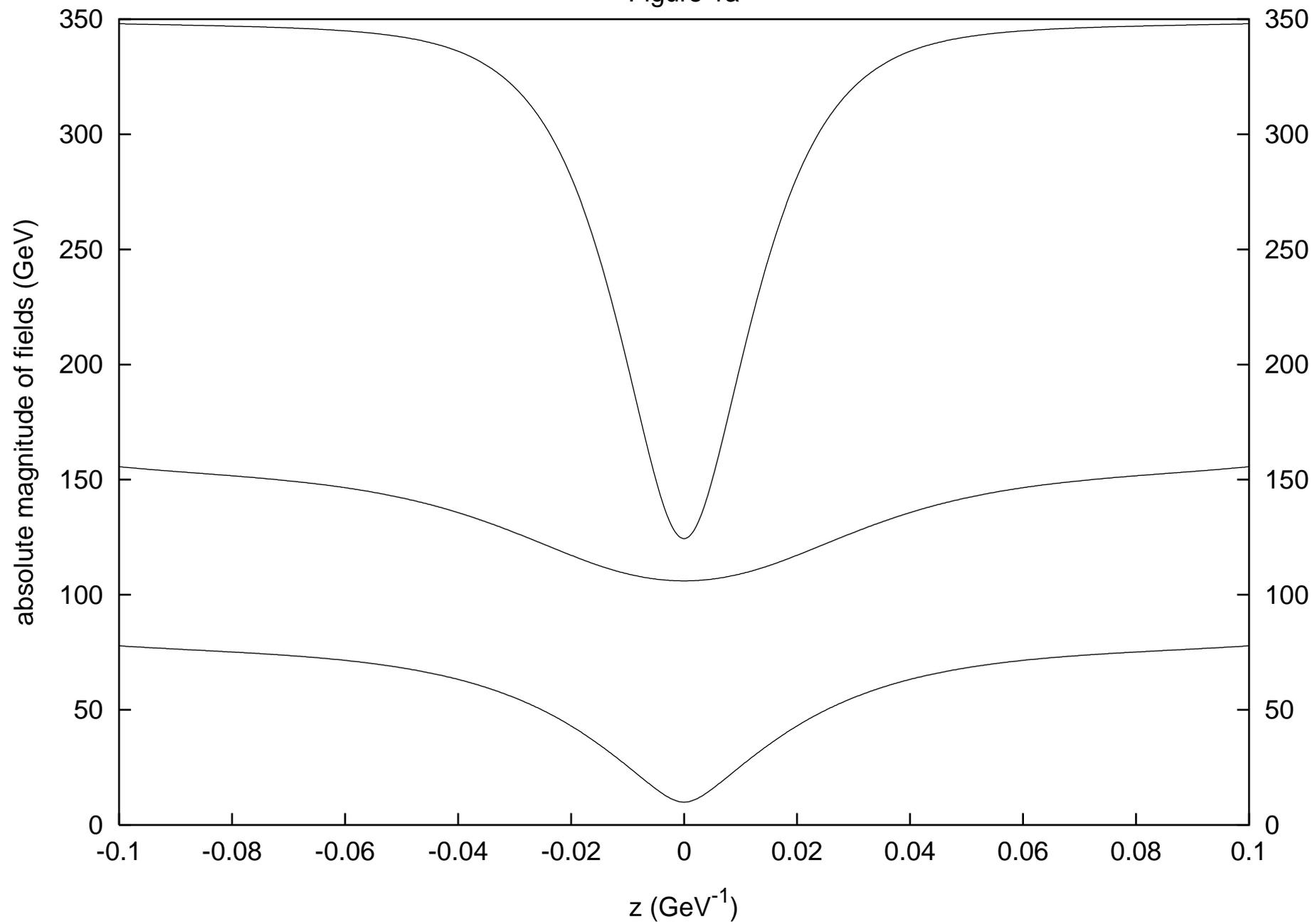
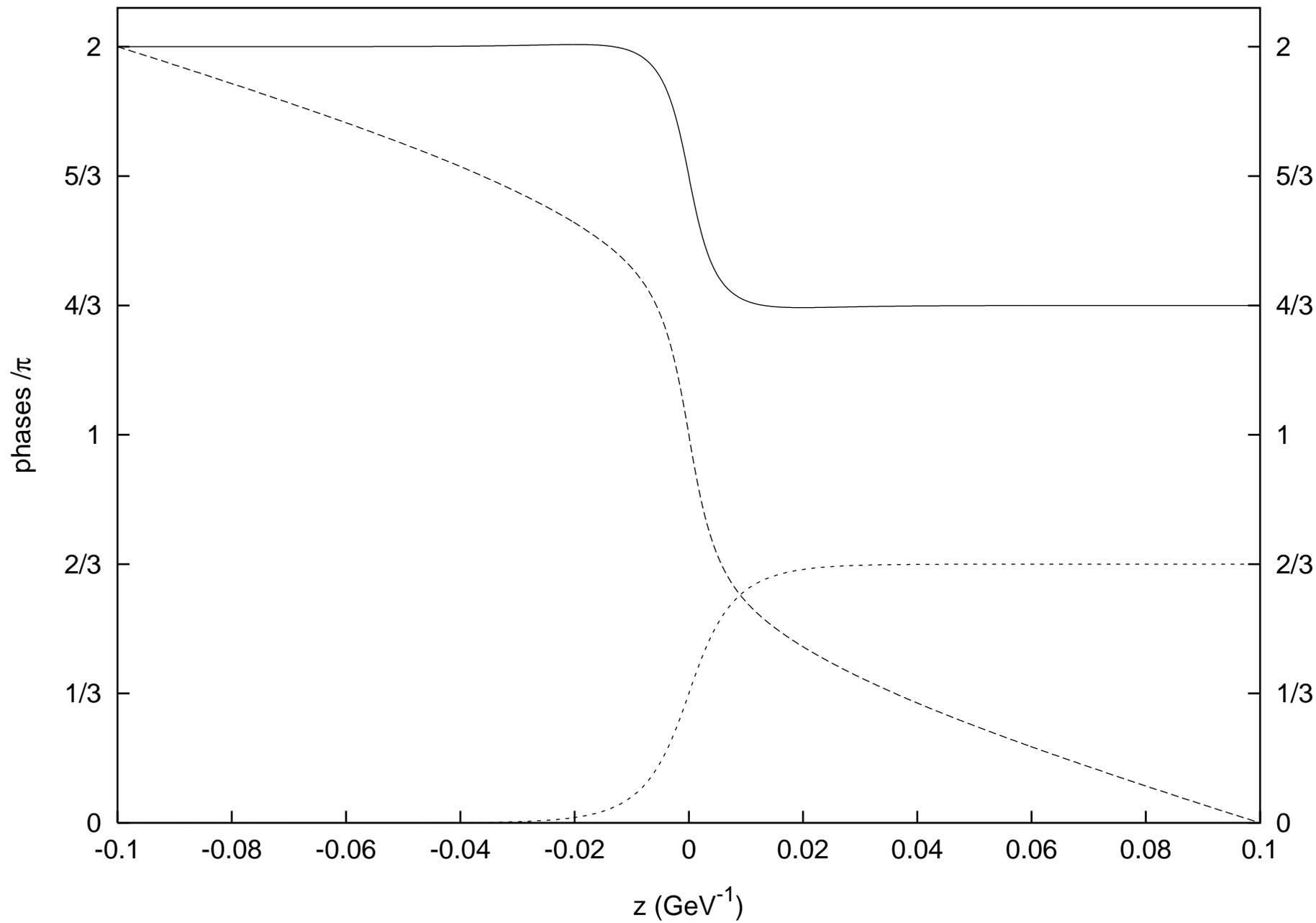
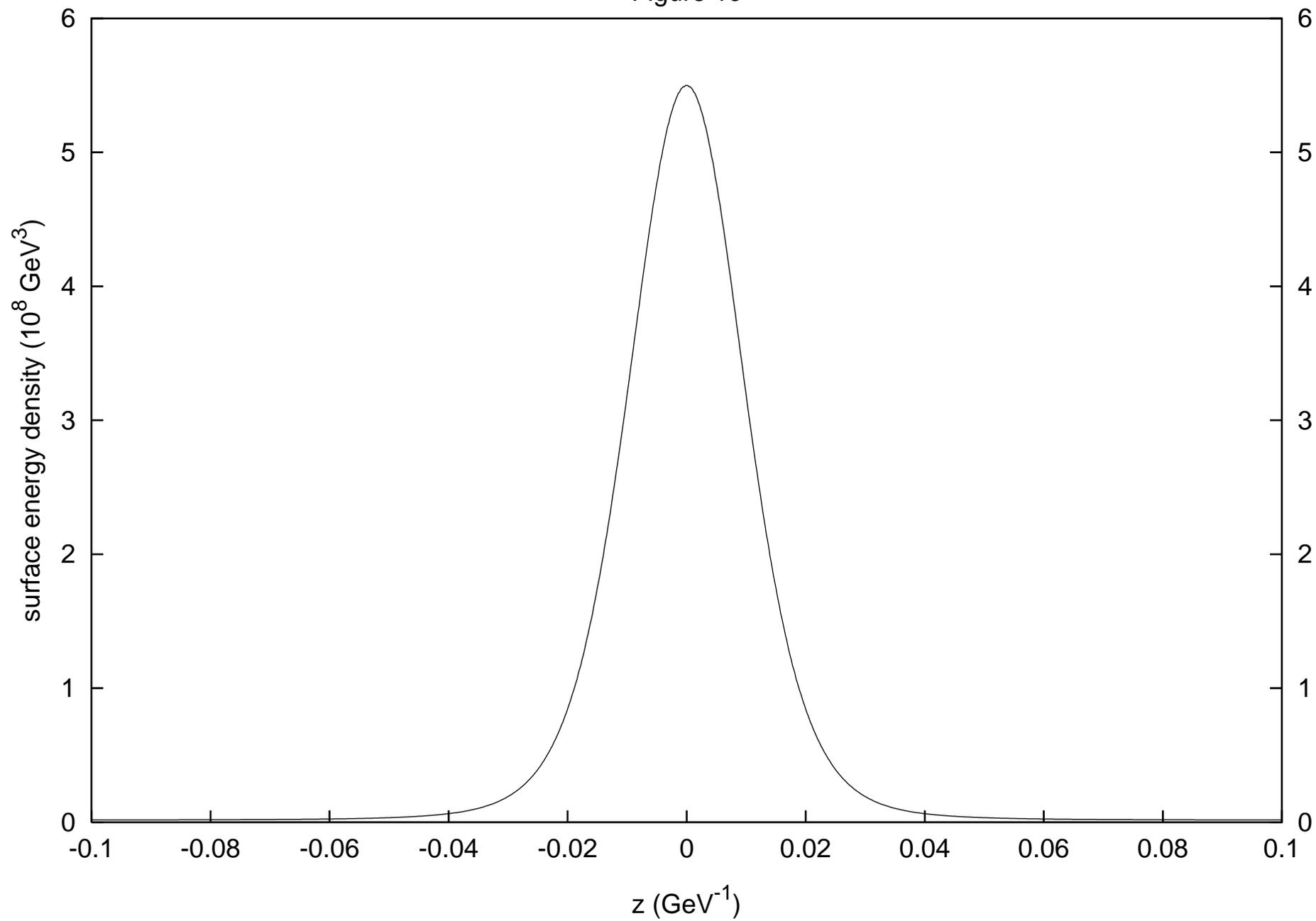
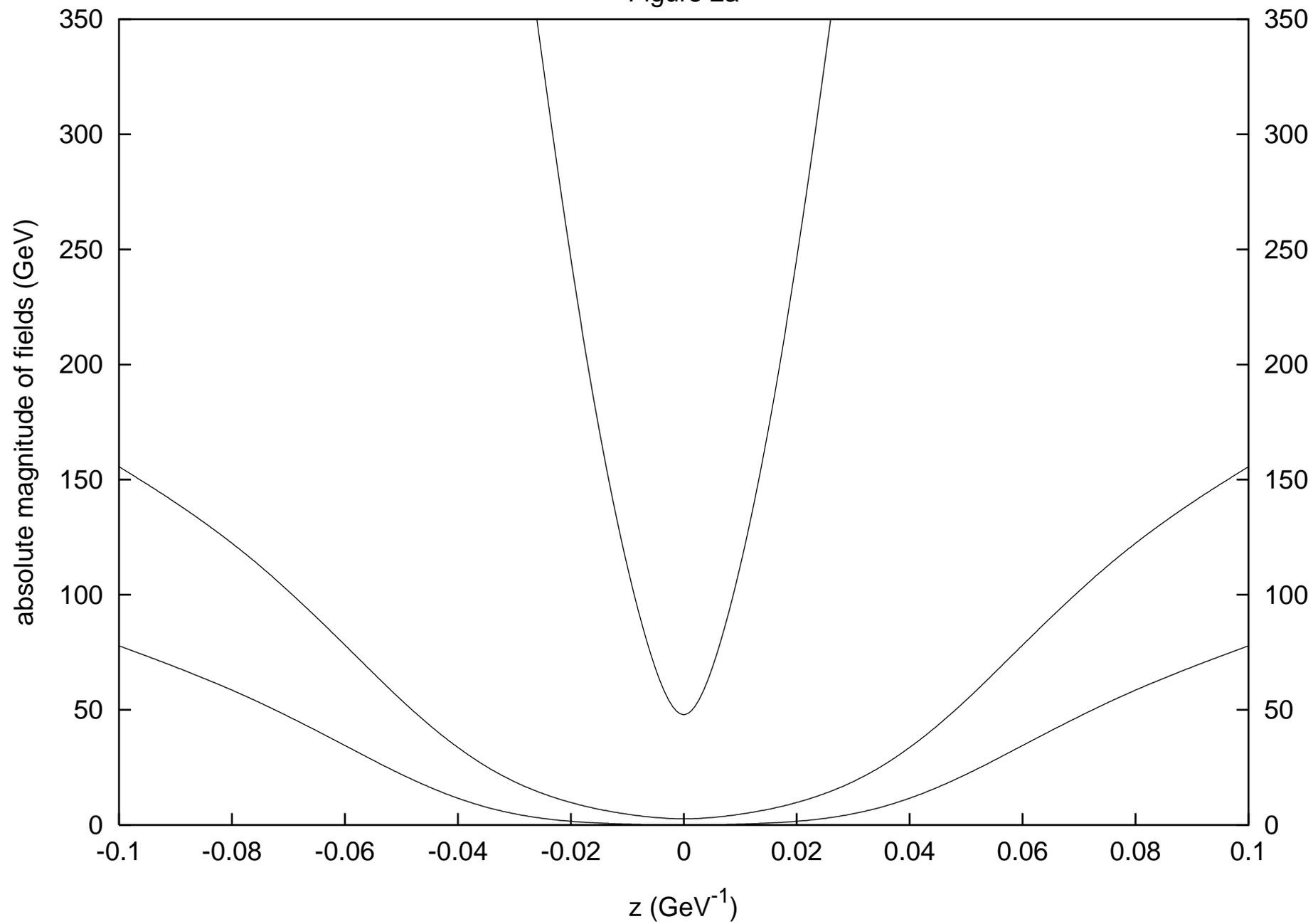
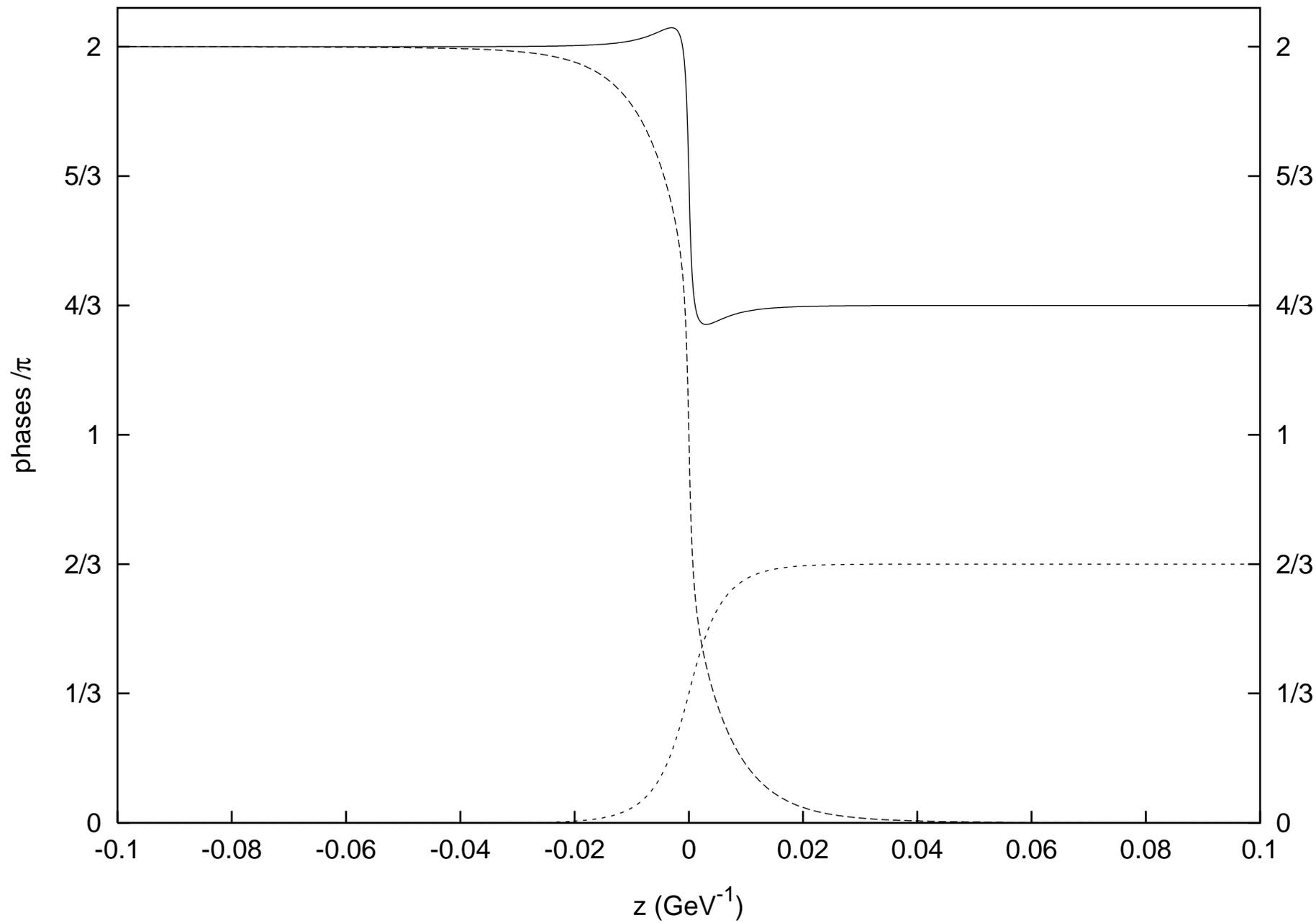
\end{document}